%% file: honma.tex
\documentclass{evn2004}
\usepackage{txfonts}
\usepackage{graphicx}
\begin{document}
\include{page}
   \title{Astrometry of W49N -- OH43.8-0.1 H$_2$O maser pair with VERA}

   \author{
Mareki \textsc{Honma},\inst{1,2}
Takeshi \textsc{Bushimata},\inst{1,3}
Yoon Kyung \textsc{Choi},\inst{1,4}
Takahiro \textsc{Fujii},\inst{1,5}
Tomoya \textsc{Hirota},\inst{1,2}\\
Koji \textsc{Horiai},\inst{1,6}
Hiroshi \textsc{Imai},\inst{5}
Noritomo \textsc{Inomata},\inst{5}
Jose \textsc{Ishitsuka},\inst{1}
Kenzaburo \textsc{Iwadate},\inst{1,6}\\
Takaaki \textsc{Jike},\inst{1}
Osamu \textsc{Kameya},\inst{1,2,6}
Ryuichi \textsc{Kamohara},\inst{1,5}
Yukitoshi \textsc{Kan-ya},\inst{1}\\
Noriyuki \textsc{Kawaguchi},\inst{1,2}
Hideyuki \textsc{Kobayashi},\inst{1,4}
Seisuke \textsc{Kuji},\inst{1,6}
Tomoharu \textsc{Kurayama},\inst{1,4}\\
Seiji \textsc{Manabe},\inst{1,2,6}
Takeshi \textsc{Miyaji},\inst{1}
Akiharu \textsc{Nakagawa},\inst{5}
Kouichirou \textsc{Nakashima},\inst{5}\\
Toshihiro \textsc{Omodaka},\inst{5}
Tomoaki \textsc{Oyama},\inst{1,4}
Satoshi \textsc{Sakai},\inst{1,6}
Katsuhisa \textsc{Sato},\inst{1,6}\\
Tetsuo \textsc{Sasao},\inst{1,7}
Katsunori M. \textsc{Shibata},\inst{1}
Rie \textsc{Shimizu},\inst{5}
Kasumi \textsc{Sora},\inst{5}\\
Hiroshi \textsc{Suda},\inst{1,4}
Yoshiaki \textsc{Tamura},\inst{1,2,6} and 
Kazuyoshi \textsc{Yamashita}\inst{5}
          }

   \institute{VERA Project, NAOJ, Mitaka, Tokyo 181-8588, Japan
         \and
        Graduate University for Advanced Studies, Mitaka, Tokyo 181-8588, Japan
         \and
        Space VLBI Project, NAOJ, Mitaka, Tokyo 181-8588, Japan
         \and
        Department of Astronomy, University of Tokyo, Bunkyo, Tokyo 113-8654, Japan
         \and
        Faculty of Science, Kagoshima University, Korimoto, Kagoshima, Kagoshima 890-0065, Japan
         \and
        Mizusawa Astro-Geodynamics Observatory, NAOJ, Mizusawa, Iwate 023-0861, Japan
         \and
        Ajou University, Suwon 442-749, Republic of Korea}

   \abstract{
We present the results of multi-epoch VERA observations of W49N -- OH43.8-0.1 H$_2$O maser pair.
Based on the dual-beam VLBI observation with VERA, we successfully obtained the phase-referenced maps of OH43.8-0.1 with respect to the W49N reference spot for 3 epochs with a time span of 6 months.
The maps were in good agreement with previous studies obtained with a single-beam VLBI, and were also consistent with each other with an accuracy of about 0.2 mas.
Moreover, there are systematic, rather linear displacements of maser feature positions, which may be the relative proper motions of maser features caused by the Galactic rotation as well as internal motions of individual maser features.
   }

   \titlerunning{Astrometry of W49N -- OH43.8-0.1 maser pair}
   \authorrunning{M. Honma {\it et al.}}
   \maketitle
%
%________________________________________________________________

\section{Introduction}

VERA (VLBI Exploration of Radio Astrometry, Sasao 1996; Kobayashi et
al. 2003 and references therein) is a VLBI array dedicated to phase
referencing astrometry.  Based on simultaneous dual-beam observations
of target and reference sources, VERA can effectively cancel out the
tropospheric fluctuation, and will enable us to measure proper motions
and parallaxes of Galactic H$_2$O and SiO maser sources with 10
$\mu$as-level accuracy.  Based on such high-precision astrometry, one
can establish 3-D structure and dynamics of the Milky Way with
unprecedentedly high accuracy (for detailed scientific targets, see
Honma et al. 2000).  Since the completion of all four stations in 2002,
we have been conducting test observations to evaluate its capability of
phase-referencing as well as astrometric performance.  Recently VERA's
high capability of phase-referencing was demonstrated based on
observations of W49N -- OH43.8-0.1 maser pair (Honma et al. 2003), and
here we report the current status of astrometric performance evaluation
using the same pair sources.
%__________________________________________________________________

\section{Observations and Reductions}

Monitoring of the W49N -- OH43.8-0.1 H$_2$O maser pair (0$^\circ$.65
separation) has been performed with a typical interval of one months,
and here we present 3 epochs under relative good conditions on day of
year 026, 120, and 205 in 2004.  Observations for each epoch lasted for
8 hours and were done with all 4 stations of VERA in the dual-beam
mode.  A bright continuum source, TXS 1923+210, was also observed every
2 hours as a clock and bandpass calibrator.  Details of the
observations and correlation processes such as recording rate,
bandwidth, frequency resolution, and so, can be found in Honma et
al. (2003; 2004).  After the correlation processing with the Mitaka FX
correlator, visibilities of all velocity channels of W49N and
OH43.8-0.1 were phase-referenced to the W49N reference maser spot at
$V_{\rm LSR}$ of 9 km s$^{-1}$, which is one of the brightest spots,
and shows no sign of structure according to the closure phase.
Phase-referenced visibilities were Fourier transformed to synthesize
images, and the positions of the brightness peaks were determined with
respect to the reference spot.

%   \begin{figure*}
%   \centering
%   %%%\includegraphics{empty.eps}
%   %%%\includegraphics{empty.eps}
%   %%%\includegraphics{empty.eps}
%   \vspace{230pt}
%   \special{psfile=sample_fig1.ps hscale=70 vscale=70 hoffset=80 voffset=0}
%   \caption{Pre-EVN map of Andromeda, from Thomas (\cite{thomas30})
%            \label{fig:thomas}
%           }
%    \end{figure*}
%
%______________________________________________________________

\section{Results}
\subsection{OH43.8-0.1 map}
We have obtained phase-referenced maser maps for both W49N and
OH43.8-0.1 (for the results for W49N, see Honma et al. 2004).  Here we
show in figure 1 the OH43.8-0.1 maser map which are phase-referenced
with respect to the W49N reference spot at $V_{\rm LSR}$=9 km s$^{-1}$.
The spot positions in figure 1 represent the sum of the position offset
of OH43.8-0.1 maser spot from its tracking center and that of the
reference spot from W49N tracking center.  In total, we have identified
38 maser spots in OH43.8-0.1 with velocity range from 30 km s$^{-1}$ to
50 km s$^{-1}$.  The spot distribution in figure 1 is in good agreement
with that of previous studies (e.g., Downes et al. 1979), showing a
ring-like spot distribution.  The radial velocity structure also
agreements with that of the previous study.

%%% figure 1
   \begin{figure}
   \centering
   \includegraphics[angle=-90,width=8.7cm]{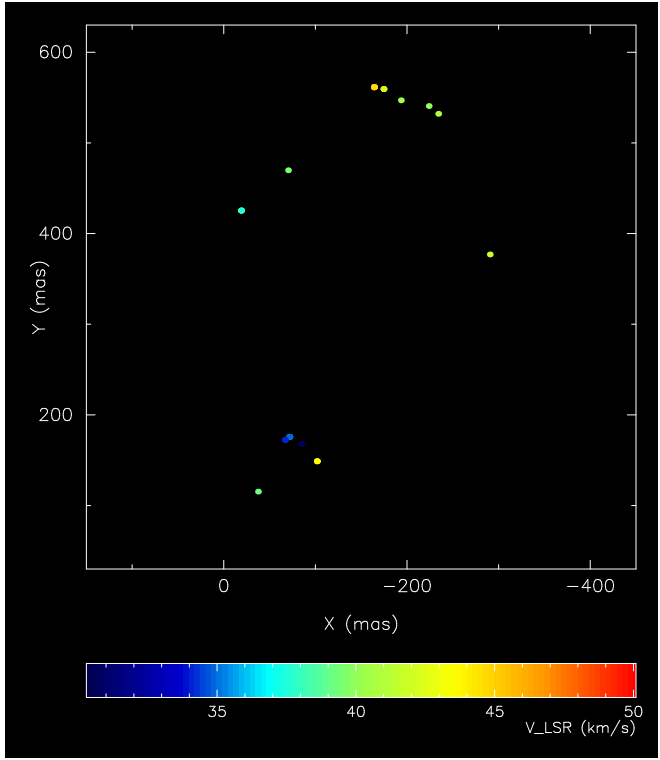}
      \caption{Phase referenced map of OH43.8-0.1 with respect to W49N
              reference maser spot.  }
         \label{FigVibStab}
   \end{figure}
%%%

\subsection{Maser positions at 3 epochs}

In order to evaluate astrometric performance of VERA, here we compare 3
epoch observations of W49N -- OH43.8-0.1 pair which have a time span of
about 6 months.  Figure 2 shows a superposed map for one of OH43.8-0.1
maser features.  The maser spots in figure 2 have $V_{\rm LSR}$ of
$\sim$ 43 km s$^{-1}$.  The spot positions for 3 epochs agree with each
other with an accuracy of 0.2 mas.  Thus, most conservatively one can
conclude that current VERA system has a positional accuracy of around
0.2 mas, which is already enough to perform astrometry within 1 kpc
from the Sun.

\section{Discussion}

In addition to position consistency within 0.2 mas, in figure 2 there
are notable displacements of the maser feature.  The position shifts
are systematic from south to north, indicating a linear motion with
almost constant velocity.  Therefore, it is possible that we have
detected maser spot motions of OH43.8-0.1 with respect to W49N
reference spot.  Maser features of OH43.8-0.1 other than that shown in
figure 2 also show this kind of systematic motion, though the direction
of the motion varies from feature to feature (which is likely to be due
to internal motions).  Comparisons of maps of OH43.8-0.1 masers that
are phase-referenced to one of OH43.8-0.1 spots also give consistent
results with figure 2, indicating that the maser feature motions are
likely to be real.

In order to obtain stronger conclusions, it is necessary to obtain more
data for longer time span and also to improve the delay calculation
model for Mitaka FX correlator, which is crucial to high-precision VLBI
astrometry, and then we will be hopefully able to perform maser
astrometry in the Galactic scale.

%%% figure 2
   \begin{figure}
   \centering
   \includegraphics[angle=-90,width=8.7cm]{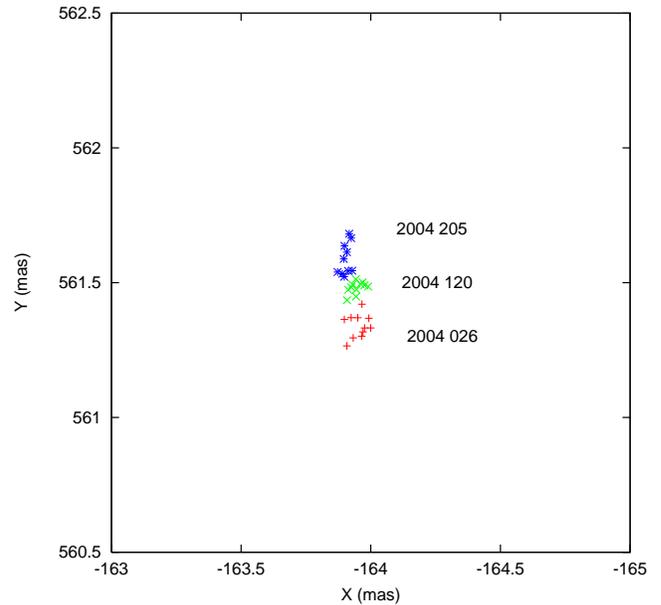}
      \caption{A superposition of OH43.8-0.1 maser feature observed at
      different epochs.  }
         \label{Fig2}
   \end{figure}
%%%

%\section{Conclusions}

\begin{acknowledgements}
One of the authors (MH) acknowledges financial support from Inamori
Foundation and that from the Ministry of Education, Culture, Sports,
Science and Technology (No.16740120).

\end{acknowledgements}

\end{document}

%% file: page.tex
\setcounter{page}{203}